# Hot Biexcitons Driven by Extreme Optical Confinement

Xinyi Wang[1], Kaushik Kudtarkar[2], Wenjing Wu[3,4], Yunjo Jeong[2], Yuxuan Cosmi Lin[1,5], Xiaofeng Qian[1,5,6], Junichiro Kono[3,7,8], Shengxi Huang[3,7], and Shoufeng Lan[1,2,5*]

[1]Department of Materials Science and Engineering, Texas A&M University, College Station, TX, 77843, USA.

[2]Department of Mechanical Engineering, Texas A&M University, College Station, TX, 77843, USA.

[3]Department of Electrical and Computer Engineering, Rice University, Houston, TX 77005 USA.

[4]Applied Physics Graduate Program, Smalley–Curl Institute, Rice University, Houston, TX 77005, USA.

[5]Department of Electrical and Computer Engineering, Texas A&M University, College Station, TX 77843, USA.

[6]Department of Physics and Astronomy, Texas A&M University, College Station, TX 77843, USA.

[7]Department of Materials Science and NanoEngineering, Rice University, Houston, TX 77005, USA.

[8]Department of Physics and Astronomy, Rice University, Houston, TX 77005, USA.

Corresponding author: shoufeng@tamu.edu

**Abstract**

A powerful means to understanding condensed matter that possesses a multi-constituent, non-isolated, and complex nature, with a preeminent example being two-dimensional (2D) materials, is studying many-body interactions. However, experimentally observing high-order many-body interactions is a daunting task due to its heavy reliance on the abundance of low-order complexes. Here, we report the observation of four-body hot biexcitons in an energetically unfavorable bilayer of tungsten disulfide ($WS_2$) through creating extreme optical confinement. Specifically, we integrate a non-radiative bound state in the continuum (BIC) into a photonic crystal (PhC) defect cavity, forming a quasi-three-dimensional (q-3D) but open confinement for photons at the driving frequency. The extremely confined photons in both reciprocal and physical spaces then excite inherently unproductive two-body hot excitons situated slightly above the indirect bandgap so efficiently that they form overwhelmed higher-order four-body hot biexcitons. Distinctively, these hot biexcitons exhibit substantial valley polarization and coherence at room temperature, which we attribute to the topological nature of BICs and the associated q-3D confinement with an orbital angular momentum. Besides achieving room-temperature biexcitons, the q-3D confinement could be valuable for higher-order interactions, such as triexcitons, and many other many-body phenomena, including Bose-Einstein condensation.

## Introduction

Controlling many-body excitonic interactions at room temperature remains a critical challenge in advancing next-generation optoelectronics using atomically thin semiconductors[1-3]. Among these materials, 2D transition-metal dichalcogenides (TMDs) stand out for their fascinating valley physics[4]. More importantly, their quantum confinement with reduced dielectric screening provides an ideal platform for the simplest but most profound two-body interactions via excitons (Xs). These two-body Xs are bound electrons and holes that can recombine and emit light efficiently, for example, for photoluminescence (PL) under moderate optical pumping. Meanwhile, higher-order four-body biexcitons (XXs) require simultaneous spatial and temporal overlap of two Xs[5]. As such, the XX generation is intrinsically nonlinear and strongly dependent on the density of lower-order Xs[1,6]. This density requirement becomes especially difficult to meet at room temperature, where phonon-mediated scattering rapidly depletes the X population and shortens lifetimes[7,8]. Consequently, biexciton emission in TMDs has been resolved predominantly at low temperature[9], whereas at room temperature it is generally weaker, broader, and more difficult to identify unambiguously even in energetically favorable monolayers[10].

Different from monolayers, Xs in bilayers are called hot excitons (AXs) because they are situated slightly above an energetically unfavorable indirect bandgap. This unfavorable energy condition makes it even more difficult, if not impossible, to achieve a high density of AXs at room temperature. Additionally, because of the inversion symmetry that goes against the spin-valley locking rule in monolayers, valley polarization in bilayers is thought to be vanishingly weak. However, recent studies suggest that intrinsic crystal symmetries, such as inversion symmetry and stacking order, play a more decisive role in valley polarization than extrinsic processes alone[11]. For instance, bilayer $WS_2$ exhibits robust and more significant valley polarization compared to monolayers due to spin-valley-layer coupling and symmetry-protected suppression of interlayer hopping, enabling strong valley readout from excitonic emissions[11,12]. Therefore, a fascinating and fundamental quest is to seek AX-induced four-body hot biexcitons (AXXs) in bilayers[6], yet this remains elusive and technically challenging with conventional technologies at room temperature.

In this work, we experimentally demonstrate room-temperature AXXs in bilayer $WS_2$ with substantial valley polarization and coherence by creating an extreme q-3D confinement at driving frequency. Specifically, the q-3D confinement integrates two complementary mechanisms: vertically, a BIC offers symmetry-protected radiation suppression to form an effective confinement in k-space[13-16]; laterally, a photonic bandgap generated by tuning the geometry of the PhC leads to an in-plane cavity with a confinement in real-space[17-20]. Such combined reciprocal and physical q-3D confinement dramatically reduces radiative pathways for the pumping photons, forcing them to only evanescently interact with the nearby bilayer $WS_2$ to excite AXs, despite the unfavorable indirect bandgap[19,21]. Unlike existing works, our approach focuses on the driving field with extreme q-3D confinement, addressing the critical density requirement for AXX generation by enabling AX accumulation instantaneously after the pump[22,23]. Moreover, the q-3D confinement keeps the topological polarization vortex of BIC with orbital angular momentum intact to improve the valley polarization and coherence of AXs and, very importantly, largely preserves it with AXXs.

## Results

### Creating q-3D extreme optical confinement

We established the physical design and operating principle of the q-3D confinement used throughout this work, as shown in **Fig. 1**. **Fig. 1a** presents the overall device concept: bilayer $WS_2$ is integrated with the PhC, enabling strongly confined photons to enhance the local excitation density in the bilayer region and providing access to both AX and AXX emissions. The spatial locations of the q-3D confinement region and the reference region on the same $WS_2$ flake are shown in **Fig. S1**. **Fig. 1b** further confirms this effect by comparing the PL spectra collected from the same bilayer $WS_2$ flake inside (blue) and outside (red) the q-3D confinement, where the region exhibits clearly enhanced AX and AXX peaks. The insets illustrate AX as a bound two-body pair of an electron (−) and a hole (+), and AXX as a four-body system combining two AXs.

**Figs. 1c–1e** illustrate the design that combines an out-of-plane confinement mechanism based on a non-radiative BIC with an in-plane confinement provided by a defect-type PhC heterostructure. The PhC square lattice in this design consists of two regions: the core and the cladding. The inner core is defined by a hole radius $r_1$, whereas the outer cladding uses a smaller radius $r_2$. Note that a five-period transition ring is inserted between them (not shown here), which will not compromise the underlying mechanism but will smooth the in-plane band discontinuity and reduce scattering loss at the heterointerface. Bilayer $WS_2$ is placed on top of the PhC structure to access AX and AXX emissions driven by the q-3D confinement. The fabricated PhC geometry agrees well with the design, as confirmed by SEM imaging (**Fig. S2**).

**Figs. 1f–1h** provide the band structure basis of this design. We calculate the photonic band structures of the two PhC components and find that the inner-layer PhC with radius $r_1$ (**Fig. 1f**) supports a photonic band (green) with a Γ-point BIC (orange dot) at the target pumping energy (dashed line). The inset of **Fig. 1f** is the first Brillouin zone of the photonic band structure. In contrast, the outer-layer PhC with radius $r_2$ shifts the relevant bands downward (flat-head arrow) and opens a bandgap (double-head arrow) within the same spectral range (**Fig. 1g**), thereby suppressing the lateral propagation of pumping photons. When the two band structures are overlaid (**Fig. 1h**), the Γ-point BIC mode lies within the bandgap region of the outer PhC (**Fig. 1h**), indicating simultaneous suppression of out-of-plane radiation through the BIC and in-plane leakage through the heterostructure-defined mode gap, consistent with the intended q-3D confinement.

**Realizing topological polarization vortices**

The q-3D confinement supports a Γ-point symmetry-protected BIC at 532 nm. As shown in the simulated and measured angle-resolved reflection spectra in the **Fig. S4**, the resonance linewidth progressively narrows toward normal incidence, confirming the presence of a BIC at the Γ point near 532 nm. The corresponding Q-factor evolution extracted from the band structure is shown in **Fig. S5**, which shows a sharp increase toward the Γ point, consistent with the symmetry-protected BIC.

We then perform polarization-resolved far-field analysis under left-circularly polarized (LCP) excitation, which enables direct visualization of the phase winding and polarization textures, thereby revealing the topological polarization vortex associated with the BIC mode. The results are separated into the cross-polarization channel (opposite handedness relative to the incident LCP) and the co-polarization channel (same handedness as the incident LCP), shown in the top row (**Figs. 2a–2d**) and bottom row (**Figs. 2e–2h**), respectively. For each channel, we present the reconstructed phase map and local polarization orientation (**Figs. 2a** and **2e**; color encodes phase and arrows indicate the local polarization direction), the corresponding interference pattern (**Figs. 2b** and **2f**), the simulated k-space intensity profile (**Figs. 2c** and **2g**), and the experimentally measured k-space intensity profile (**Figs. 2d** and **2h**).

In the cross-polarization channel (**Figs. 2a–2d**), the phase distribution around Γ exhibits a clear singularity, while the polarization-orientation field undergoes two full rotations upon encircling the center, consistent with a second-order polarization vortex. Quantitatively, the accumulated phase along a closed loop around Γ approaches $-4\pi$, corresponding to a topological charge of $q = -2$[14,15,24].

$$q = \frac{1}{2\pi}\oint \nabla\phi \cdot dl = -2$$

Consistent with this q = −2 winding, the interference map shows a clear two-armed spiral texture (second column). The simulated cross-polarized k-space intensity profile exhibits a pronounced four-armed pattern forming a quadrilateral-like envelope (third column). The experimentally measured k-space pattern shows a similar four-armed, quadrilateral-like intensity distribution (fourth column), confirming good agreement between the simulated and measured topological far-field patterns.

By contrast, in the co-polarization channel (**Figs. 2e–2h**), the reconstructed phase and polarization-orientation textures become uniform. The interference pattern evolves into nearly concentric rings, and the k-space intensity distribution correspondingly loses the four-armed structure, reverting instead to a Gaussian-like profile associated with the incident beam. In this channel, no distinct topological feature is observed.

Here, the sign of the measured topological charge follows the handedness of the incident circular polarization. When the input is switched from LCP to RCP, the charge reverses from q = −2 to q = +2, together with the handedness of the spiral texture. Consistently, upon reversing the incident helicity in the experiment, the emitted k-space pattern changes from clockwise to counterclockwise rotation as the analyzer is rotated, in agreement with the simulated sign reversal of the topological charge.

The vortex measurements performed on the pure-BIC PhC device and across the entire q-3D device (**Fig. S9**) exhibit the same topological characteristics. This result indicates that the observed vortex property of the q-3D device originates from the Γ-point BIC supported by the core region, while the surrounding heterostructure mainly provides the additional in-plane confinement required for the q-3D structure.

In the present system, resolving this vortex is important because the BIC polarization singularity can, through spin-orbit or geometric-phase coupling, generate a structured pump field with nontrivial angular momentum[25-27]. Such a field provides a plausible route for coupling the q-3D-confined photonic mode to the valley degree of freedom in bilayer $WS_2$. Although the detailed microscopic mechanism remains beyond the scope of the present work, this topology-mediated excitation may contribute to the enhanced valley polarization and coherence observed in AX and AXX emissions discussed below.

### Achieving AXXs and experimental verification

After establishing q-3D confinement and the topological property associated with orbital angular momentum, we transfer bilayer $WS_2$ prepared by a gold-assisted exfoliation method onto the device for PL characterization[31]. **Fig. 3a** and **3b** compare PL spectra measured at a pump power density of 1.85 kW/cm$^2$. We measure the spectrum with confinement at the center of the q-3D core region and acquire the spectrum without confinement from the same $WS_2$ flake outside the PhC region. Outside the PhC, $WS_2$ shows only a weak and broad emission feature near 611 nm. Under the same pump condition, the PL with q-3D confinement shows two distinct peaks centered at 611 nm (Peak 1) and 623 nm (Peak 2), with the overall emission intensity strongly enhanced, consistent with pump-assisted excitation enhancement via the q-3D design.

Peak 1 corresponds to neutral AX emission based on its wavelength. To determine the origin of Peak 2, we perform pump-density-dependent PL measurements. As shown in **Fig. 3c**, Peak 2 near 623 nm shows a nonlinear increase with excitation power, growing significantly faster than Peak 1 and becoming dominant at high pump power densities. We extract the peak heights using Gaussian fits and plot the relationship between their intensities on a log-log scale (**Fig. 3d**). We fit the data with a power-law relation $I_{AXX} \propto I_{AX}^{\alpha}$, where $I_{AXX}$ is the intensity of AXXs, $I_{AX}$ is for AXs, and $\alpha$ is the power or the slope in log-log plots. Such a power law is widely used to characterize excitonic behavior; for example, $\alpha$ close to 1 indicates a dark exciton[32]. Here, we obtain $\alpha = 1.87$, which is close to 2. Similar nonlinear behavior was observed in previous studies of biexcitons at low temperature, indicating a four-body nature[5,9]. Our observed $\alpha$ is consistent with this nonlinear behavior and thus supports assigning Peak 2 to AXX emission at room temperature for the first time[1,6].

### Room-temperature valley polarization and coherence of AXXs

To further probe excitonic states driven by the q-3D confinement, we measure the valley polarization and valley coherence in bilayer $WS_2$ both inside and outside the q-3D regime. **Figs. 4a** and **4d** show circularly polarized PL spectra measured under 532-nm LCP excitation, with and without confinement regions, respectively. Outside the q-3D confinement, the circularly polarized PL spectra show little discernible circular dichroism or valley polarization, whereas inside the confinement region, both the AX and AXX peaks exhibit clear circular dichroism. We define the degree of valley polarization as

$$P_V = \frac{I_{\sigma^+} - I_{\sigma^-}}{I_{\sigma^+} + I_{\sigma^-}}$$

Here, $P_V$ denotes the valley polarization degree, and $I_{\sigma^+}$and $I_{\sigma^-}$represent the PL intensities measured in the $\sigma^+$and $\sigma^-$ channels, respectively. The measured $P_V$ is approximately 3%.

In addition to valley polarization, we investigated the valley coherence for bilayer $WS_2$ under linearly polarized excitation. A linearly polarized field can be viewed as a coherent superposition of two counterrotating circularly polarized components with a relative phase, where the phase determines the linear polarization axis. In TMDs, such excitation can access valley coherence and thus yield linearly polarized emission components[3,4,33,34]. **Figs. 4b** and **4e** show the polarization-resolved PL spectra measured under horizontal linear excitation ($H_{in}$), with analysis of the co-polarized ($H_{in}$, $H_{out}$) and cross-polarized ($H_{in}$, $V_{out}$) output components inside and outside the q-3D confinement. Inside the q-3D confinement, the co-polarized channel exhibits a clear enhancement over the cross-polarized channel across the AX/AXX emission band, indicating a pronounced linear polarization contrast. We define the degree of linear polarization as:

$$P_L = \frac{I_{HH} - I_{HV}}{I_{HH} + I_{HV}}$$

Here, $P_L$ denotes the degree of linear polarization, and $I_{HH}$ and $I_{HV}$ represent the PL intensities measured in the co-linear ($HH$) and cross-linear ($HV$) polarization channels, respectively. Experimentally, $P_L$ is measured to be 0.24 for AX and 0.21 for AXX. These results indicate that the valley coherence of AXX is slightly reduced compared with that of AX but remains largely preserved.

Outside the confinement, the corresponding linear polarization contrast is not observable. Finally, **Figs. 4c** and **4f** plot the peak intensity as a function of the analyzer angle in polar coordinates. For the in-confinement region (**Fig. 4c**), both AX and AXX show a characteristic two-lobe dependence on the output polarization angle (**Fig. 4c**), while the AXX polar profile appears noticeably broader than that of AX. In contrast, the out-of-confinement AXX signal is nearly circular in the polar plot (**Fig. 4f**), indicating near-zero valley coherence.

## Conclusion

Our results show that extreme q-3D confinement at the driving frequency can overcome a key limitation of room-temperature many-body physics: the difficulty of building up a sufficiently dense lower-order many-body population in condensed matter. Particularly, in bilayer $WS_2$, AXs lie above the indirect bandgap and are therefore difficult to populate and sustain under ambient conditions. Against this unfavorable background, the spectrally resolved AXX emission observed under q-3D confinement indicates that the confinement does more than enhance AX intensity; it also improves the excitation process, making higher-order excitonic channels accessible even at room temperature. More broadly, this result suggests that engineering the driving field can provide an effective route to accessing excitonic complexes even in multilayer systems whose band structures are not intrinsically favorable for their formation. The valley-resolved measurements further suggest that the q-3D confinement affects not only the exciton population but also the valley selectivity. Within the q-3D confinement, both AX and AXX exhibit stronger valley polarization and valley coherence than in the absence of the q-3D confinement. We attribute this valley enhancement to the topological nature of BICs and the associated q-3D confinement with orbital angular momentum. This strategy of extreme q-3D confinement enables extension to higher-order excitonic complexes and other nonlinear or collective optical phenomena that remain difficult to access under ambient conditions.

## Methods

### Simulation

We calculated the photonic band structures of the PhC slab using the guided-mode expansion method implemented in the open-source Legume package[35]. We simulated the phase maps, interference patterns, and k-space intensity distributions in **Fig. 2** using the finite-difference time-domain method in Tidy3D. We computed the reflection spectrum of the PhC core region using COMSOL Multiphysics in the frequency domain with Floquet periodic boundary conditions. Unless otherwise specified, we performed all optical simulations without including the bilayer $WS_2$. Additional details, material parameters, boundary conditions, source configuration, and numerical settings are provided in the Supplementary Information.

### Optical device fabrication

We deposited stoichiometric $Si_3N_4$ films on Si substrates by plasma-enhanced chemical vapor deposition (PECVD). Then we defined PhC patterns in ZEP520A resist using electron-beam lithography (EBL) and transferred them into the $Si_3N_4$ layer by inductively coupled plasma reactive-ion etching (ICP-RIE) with an $SF_6$/$CHF_3$/He gas mixture. After etching, we removed the residual resist in heated N-methyl-2-pyrrolidone (Remover PG) and etched the underlying

Si in 30 wt% KOH to fully release the patterned membrane. We thoroughly rinsed and dried the suspended devices before bilayer $WS_2$ integration. **Fig. S3** illustrates the detailed fabrication process flow.

**$WS_2$ preparation and integration**

We prepared bilayer $WS_2$ flakes using the Au-assisted exfoliation method[31]. We deposited a 50-nm-thick Au film onto the exposed $WS_2$ surface on adhesive tape by electron-beam evaporation (EBE). Then we delaminated the Au/$WS_2$ stack using thermal release tape (TRT). Owing to the strong Au-S interaction, large-area thin $WS_2$ flakes were preferentially retained on the Au film. We transferred the Au-supported flakes onto an oxygen-plasma-cleaned Si/$SiO_2$ substrate, heated the sample to release the TRT, and removed the Au film using KI/$I_2$ etchant. After rinsing and drying, we identify bilayer flakes under an optical microscope.

For integration onto $Si_3N_4$ photonic structures, we prepared a PDMS-supported PPC transfer stamp by spin-coating a PPC layer onto a PDMS stamp. We then brought the PDMS/PPC stamp into conformal contact with the target $WS_2$ flake, enabling pickup via adhesion to the PPC layer. The flake was then aligned to the target $Si_3N_4$ PhC using a precision micro-alignment stage and brought into contact. We mildly heated the sample to soften the PPC and facilitate flake release, thereby completing the transfer onto the patterned $Si_3N_4$ substrate. We removed the residual PPC using chloroform. Finally, we applied a thin PMMA capping layer on top of $WS_2$ to protect the flake during subsequent handling and measurements.

**Optical Characterization**

Micro-photoluminescence (µ-PL) measurements were performed under room temperature using a home-built confocal setup (see **Fig. S8** for the optical layout). We focused a 532 nm femtosecond laser onto the device through an objective lens (Nikon 40× NA = 0.75). The calibration determines the excitation spot size and the power delivered to the sample. PL signal was collected in a backscattering geometry through the same objective lens and was analyzed with a SpectraPro HRS-300 spectrometer equipped with a PIXIS 400 CCD camera. Unless otherwise specified, we acquired spectra with a 60-second integration time.

We performed valley-resolved PL measurements using the same experimental setup, incorporating a quarter-wave plate and a linear polarizer into the excitation and detection optical paths, respectively, to project the excitation and emission light onto a selected circular polarization basis. The definitions of the polarization quantities discussed in this paper are provided in the main text.

**Data and code availability**

All data that support the findings of this study are included in the article and its supplemental information.

**Acknowledgements**

The authors gratefully acknowledge the funding support provided by Texas A&M University, United States National Science Foundation (Grant No. 2348611 and 2348610), and Sandia National Laboratories (Grant No. PO 2543653/1923579) from the United States Department of Energy (DOE). The nanofabrication was conducted in the Texas A&M University AggieFab Nanofabrication Facility (RRID: SCR_023639), which is supported by the Texas A&M Engineering Experiment Station and Texas A&M University. X.Q. acknowledges the support from the United States National Science Foundation under Grant No. DMR-2103842.

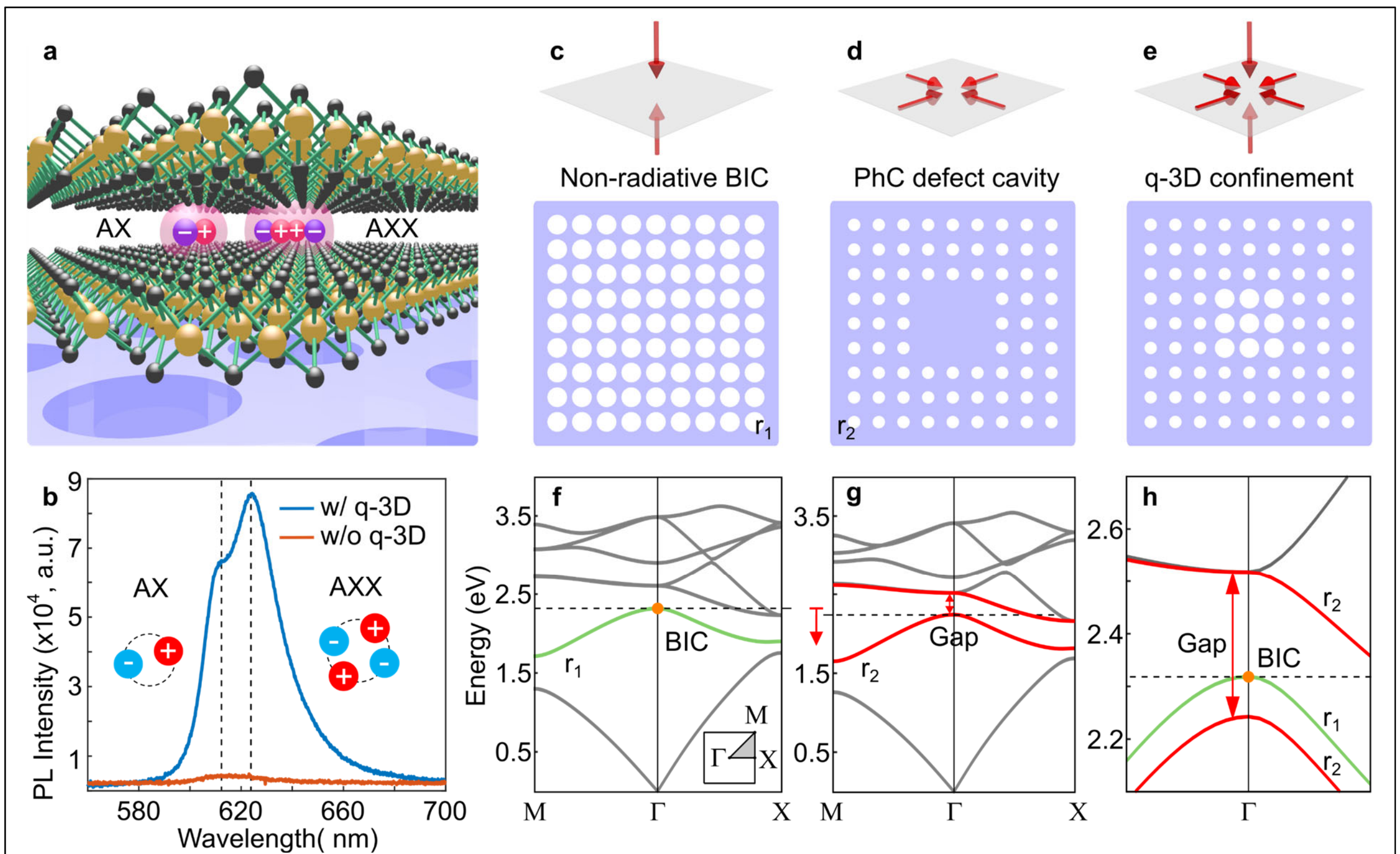

**Fig. 1: AXXs in atomic bilayers driven by q-3D confinement. a**, Schematic of two-body hot exciton (AX) and four-body hot biexciton (AXX) in bilayer $WS_2$ driven by q-3D confinement at the pumping wavelength. **b**, Room-temperature PL spectra with the q-3D confinement (blue) show not only a dramatic enhancement but also a significant AXX compared to that without the q-3D confinement (red) under the identical excitation at 532 nm. Dashed lines mark the AX (left) and AXX (right) emission wavelengths with insets showing the two- and four-body nature accordingly. **c-e**, Design principle to combine a non-radiative BIC as an out-of-plane confinement (**c**) with a PhC defect cavity for the in-plane confinement (**d**) for ultimately creating an extreme q-3D confinement (**e**). Top: conceptual sketches of confinement; red arrows indicate the directions along which light is confined. Bottom: corresponding planar layouts of lattices with different radii ($r_1$ and $r_2$) while keeping the pitch intact (not to scale). **f-h**, Simulated photonic band structures for a $Si_3N_4$ PhC with a square lattice. **f**, A symmetry-protected BIC (orange) is obtained at the Γ-point in a transverse electric or TE-like band (green) with the first Brillouin zone shown in the inset. **g**, By slightly reducing the radius to $r_2$, the TE-like band shifts to a lower energy level, indicated by a single-headed arrow. Additionally, the lattices open a semi-bandgap at the Γ-point (double-head arrow) with the two bands at the edge highlighted in red, thereby creating a PhC defect cavity in **d**. **h**, By arranging inner- and outer-layer of lattices with $r_1$ and $r_2$, the combined band structure shows that the BIC lies within the photonic bandgap, forming the q-3D confinement at the pumping wavelength (dashed line).

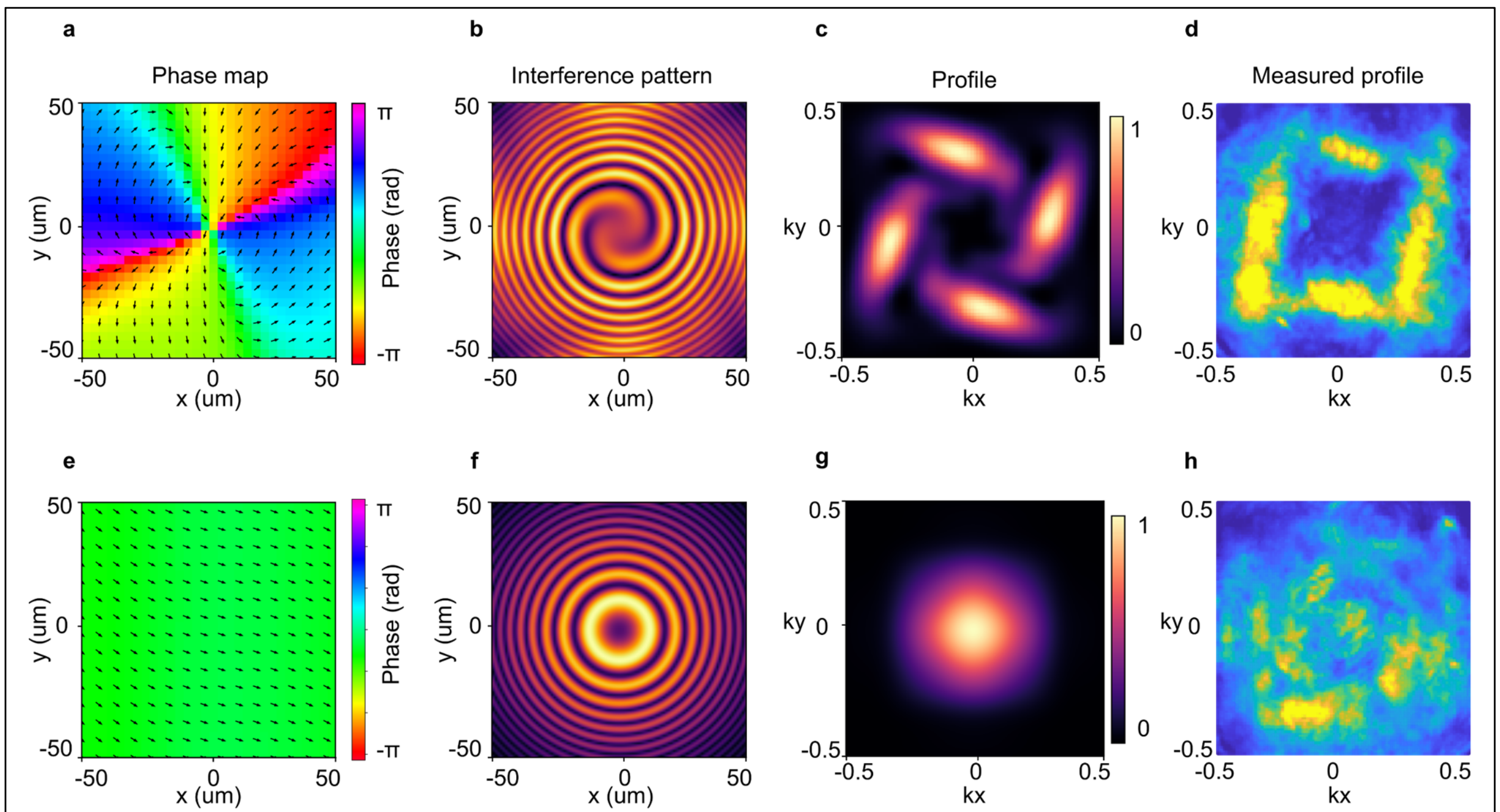

**Fig. 2: Topological polarization vortices observed in BICs. a-d**, Simulation and experimental results of vortices in a cross-polarization scenario with left circular polarization (LCP) at the input and right circular polarization (RCP) as the output. **a**, Reconstructed phase mapping (color) overlaid with the local polarization field (arrows), showing a vortex-like singularity at the center with a 4π phase change in one circle. **b**, Corresponding interference pattern, exhibiting a spiral feature consistent with the vortex phase winding. **c**, Simulated k-space intensity profile of the output field, forming a quadrilateral-like pattern composed of four distinct arms. The combined phase winding, polarization texture, and channel-dependent k-space intensity patterns confirm the topological nature of the BIC, with a winding number (topological charge) of $q = -2$. **d**, Experimentally measured k-space output intensity with cross-polarization shows a similar four-arm quadrilateral distribution, validating the polarization vortices predicted by simulation. **e–h**, Simulation and experimental results for co-polarization (LCP input, LCP output) with phase mapping (**e**) and interference pattern (**f**) in the real-space, as well as simulated (**g**) and measured (**h**) intensity profile in the k-space. All results in **e-h** show a topological charge of 0. The results for the counterpart scenario with RCP input are not shown here but exhibit the exact opposite behaviors. Though not exhibited here, similar topological polarization vortices are observed with the q-3D confinement.

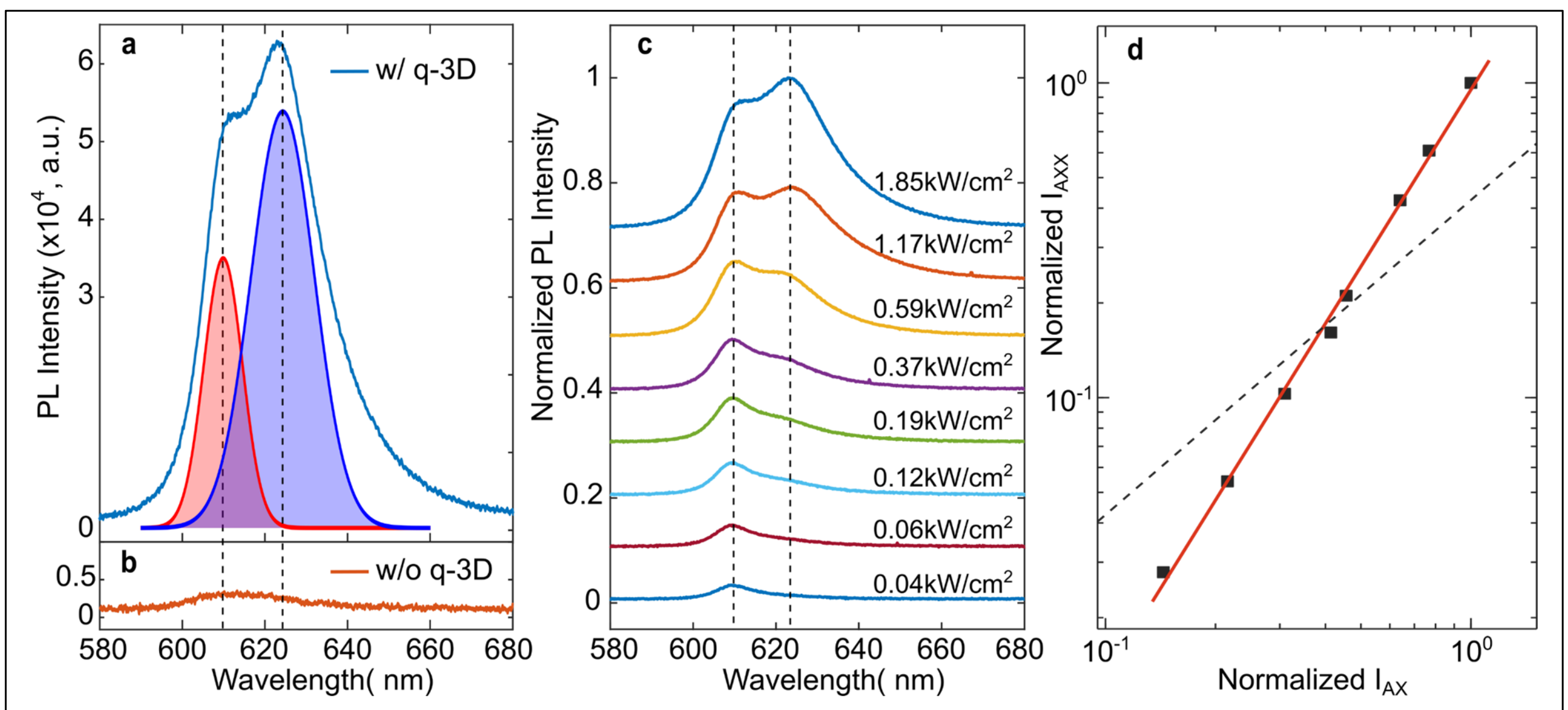

**Fig. 3: Experimental verification of AXXs. a, b**, PL spectra measured in bilayer $WS_2$ with (**a**) and without (**b**) the q-3D confinement at the pumping wavelength. The latter is obtained from the same bilayer sample but from a different area of the plain substrate. The shaded areas are the fitting, showing not only the dramatic enhancement of AXs but also the emergence of AXXs with their center wavelengths marked by dashed lines. **c**, The PL spectra (normalized and vertically offset for clarity) exhibit the rapid growth of AXXs from almost invisible at the pump power density of 0.04 kW/cm$^2$ to dominating at 1.85 kW/cm$^2$. During this process, no significant wavelength shift was observed for both AX and AXX. **d**, The log–log plot of the normalized AXX intensity, $I_{\mathrm{AXX}}$, versus the AX intensity, $I_{\mathrm{AX}}$, extracted using the fitting in **a** for each of the pump power densities in **c**. The solid red line shows a power-law fit, $I_{\mathrm{AXX}} \propto I_{\mathrm{AX}}^{\alpha}$, yielding $\alpha = 1.87$. This superlinear dependence is consistent with the assignment of the ~623 nm emission to AXX. The dashed line shows $\alpha = 1$ for comparison.

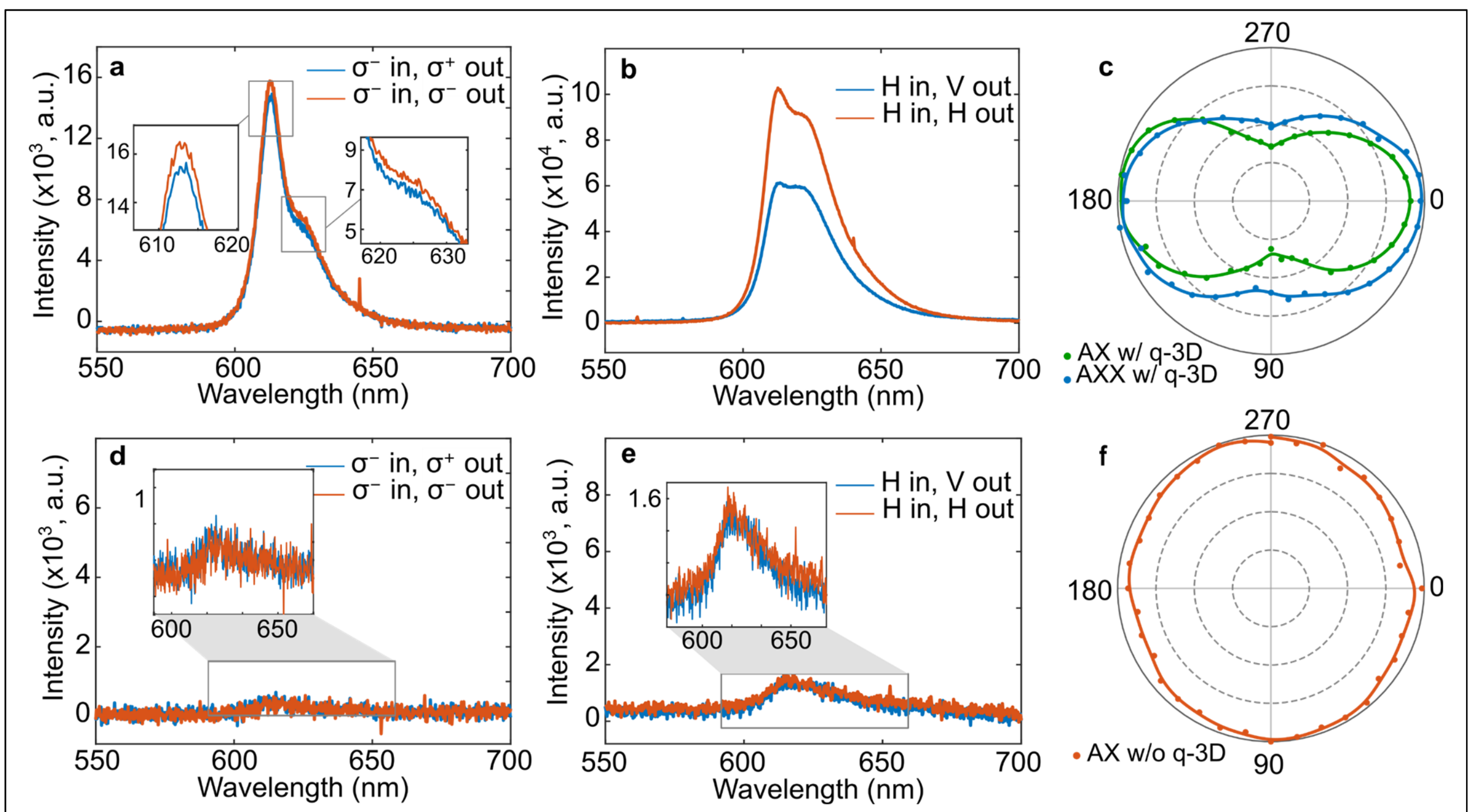

**Fig. 4**: **Valley polarization and coherence of AXXs at room temperature**. **a-c**, Measurements on bilayer $WS_2$ with the q-3D confinement at the pumping wavelength (532 nm) at room temperature. **a**, Circularly resolved PL spectra under $\sigma^-$ (LCP) excitation, analyzed with $\sigma^+$ (RCP, blue) and $\sigma^-$ (red). The insets are zoomed-in views, showing substantial valley polarization at both AX and AXX wavelengths. **b**, Linearly resolved PL spectra pumped with a fixed horizontal linear polarization (H) and analyzed with vertical (V, blue) and horizontal (red) linear components, revealing strong linear polarization anisotropy. **c**, Polar plots of the output normalized PL intensity achieved by rotating the linear analyzer. The contrast between the 0° (180°) and 90° (270°) indicates the existence of valley coherence at room temperature. Although the contrast ratio for AXXs is slightly smaller than that of AXs, the valley coherence is largely preserved. **d-f**, Corresponding measurements in the same bilayer $WS_2$ on a plain substrate without the q-3D confinement. The insets in **d** and **e** are zoomed-in views, in which no significant contrast between circular (valley) polarizations or linear polarizations. Compared to the peanut shape in **c**, the near-circular behavior in **f** also indicates a vanishing valley coherence.